\documentclass[twocolumn,twoside]{revtex4}
\usepackage{graphicx}
\usepackage{fancyhdr}
\usepackage{mathtools,multirow,gensymb,bm}
\begin{document}

\title{Exploring New Physics in $D_{(s)}^+\rightarrow\eta^{(\prime)}\bar{\ell}\nu_{\ell}$ Decays}

%

\author{Karthik Jain}\author{Barilang Mawlong}
\affiliation{School of Physics, University of Hyderabad, Hyderabad, Telangana, India, 500046}

\begin{abstract}
The observation of anomalies in the charged current $b\rightarrow c\bar{\ell}\nu_{\ell}$ transitions hints the possibility of the existence of new physics beyond the standard model. Inspired by the work done in the beauty quark sector, we explore new physics in the charm quark sector with $c \rightarrow (s,d)\bar{\ell}\nu_{\ell}$ charged current transitions. We analyze the decay modes $D_{(s)}^+ \to \eta^{(')} \bar{\ell} \nu_{\ell}$ within an effective Lagrangian framework which includes both standard model and new physics contributions. We use the available experimental measurements of the semileptonic $D$ meson decays to constrain the parameter space of new physics couplings. We then investigate the effects of new physics on four linearly independent observables, the branching fraction, forward-backward asymmetry, lepton polarization and convexity parameter.
\end{abstract}

\maketitle

\thispagestyle{fancy}


\section{Introduction}
One of the most important aspects of particle physics research today is searching for new physics (NP) beyond the standard model (SM). In recent years, measurements of the lepton flavor universality (LFU) ratios $R_{D^{(*)}}$ originating from the $b \rightarrow c \bar{\ell} \nu_{\ell}$ processes  have exhibited tensions at around $3.1 \sigma$ level from their corresponding SM predictions \cite{HFLAV:2019otj}. These $b-$physics observations have kindled interest to search for NP in charged-current transitions occuring at tree level. Motivated by the possible presence of NP in $b-$hadron decays, we explore possible NP contributions in the charm sector, involving charged-current transitions $c\rightarrow (s,d) \bar{\ell} \nu_\ell$. Charm meson decays provide a unique environment to probe flavor physics beyond the SM in the up-sector. Currently, the experimental measurements \cite{ParticleDataGroup:2020ssz} are in agreement with SM predictions \cite{Riggio2018}. These results can be used to constrain the NP parameter space. In this work, using a model-independent approach, we focus on the decay modes $D^+_{(s)}\rightarrow \eta^{(\prime)}\mu^+\nu_\mu$. Along with branching fraction, we analyse the effects of NP on observables like forward-backward asymmetry, convexity parameter and lepton polarization asymmetry. Currently, the tensor current form factor inputs are unavailable for these decay modes, so we consider only scalar and vector type of NP contributions here. We also assume the new couplings to be complex. 

\section{Theoretical Framework}
\subsection{Effective Lagrangian}
If all possible Lorentz structures are considered and only left-handed neutrinos are assumed, the effective Lagrangian for the $c\rightarrow(s,d)\bar{\ell}\nu_\ell$ transition can be written as \cite{leng2020}
\begin{eqnarray}
	\mathcal{L}_{eff} &=& -\frac{4G_F}{\sqrt{2}}V^*_{cq} [(1+C_{V_{L}}^\ell)O^\ell_{V_{L}} + C_{V_{R}}^\ell O^\ell_{V_{R}}  \nonumber \\&+&  C_{S_{L}}^\ell O^\ell_{S_{L}} + C_{S_{R}}^\ell O^\ell_{S_{R}} + C^\ell_T O^\ell_T ] + h.c.
	\end{eqnarray}
The four-fermion operators are defined as
\begin{eqnarray}
	O^\ell_{V_{L}} &=& (\bar{q}\gamma^\mu P_L c)(\bar{\nu}_\ell \gamma_\mu P_L \ell),~ O^\ell_{V_{R}} = (\bar{q}\gamma^\mu P_R c)(\bar{\nu}_\ell \gamma_\mu P_L \ell), \nonumber \\
	O^\ell_{S_{L}} &=& (\bar{q}P_L c)(\bar{\nu}_\ell P_R \ell),~ O^\ell_{S_{R}} = (\bar{q}P_R c)(\bar{\nu}_\ell P_R \ell)~, \nonumber \\
	O^\ell_T &=& (\bar{q}\sigma^{\mu\nu}P_L c)(\bar{\nu}_{\ell} \sigma_{\mu\nu}P_R \ell), 
\end{eqnarray}
and $C^\ell_i (i = V_{L}, V_{R}, S_{L}, S_{R} \ \text{and} \ T)$ are the corresponding Wilson coefficients, with $C^\ell_i = 0$ in the SM.

\subsection{Hadronic Matrix Elements}
The hadronic transition is parameterized by heavy-to-light form factors and the hadronic matrix elements for $D \to P$ transition are given by

\begin{eqnarray}
	\langle P(p_2) \vert \bar{q}\gamma^\mu c \vert D(p_1)\rangle &=& f_+(q^2)\left[(p_1+p_2)^\mu - \frac{m_D^2 - m_P^2}{q^2} q^\mu\right] \nonumber \\ & +& f_0 (q^2)\frac{m_D^2-m_P^2}{q^2}q^\mu~,
\end{eqnarray}

\begin{eqnarray}
	\langle P(p_2) \vert \bar{q} c \vert D(p_1)\rangle &=& \frac{q_\mu}{m_c-m_q}\langle P(p_2) \vert \bar{q}\gamma^\mu c \vert D(p_1)\rangle \nonumber \\ & = & \frac{m_D^2 - m_P^2}{m_c-m_q}f_0(q^2)~.
\end{eqnarray}

\subsection{$\bm{\eta-\eta^{\prime}}$ mixing}
We consider the mixing of the quark components $\eta_q = q\bar{q} = \frac{u\bar{u}+d\bar{d}}{\sqrt{2}}$ and $\eta_s = s\bar{s}$ for the pseudoscalar mesons $\eta$ and $\eta^{\prime}$, where the mixing is given by
\begin{equation}
\begin{pmatrix}
	\eta \\ \eta^{\prime}
\end{pmatrix} = \begin{pmatrix}
	\cos\theta_p & \sin\theta_p \\ -\sin\theta_p & \cos\theta_p
\end{pmatrix} \begin{pmatrix}
	q\bar{q} \\ s\bar{s}
\end{pmatrix} .
\end{equation}
We use $\theta_p = (40.1\pm 2.1\pm 0.7)\degree$ \cite{Ablikim2019} in our work.

\subsection{Form Factors}
In this work, we use the form factors obtained from light cone sum rules (LCSR) \cite{Wu2006}. The following parametrization describes a reasonable behavior of the form factors in the whole kinematically accessible region
\begin{equation}
F^i(q^2) = \frac{F^i(0)}{1 - a\frac{q^2}{m^2_D} + b\left(\frac{q^2}{m^2_D}\right)^2}~,
\end{equation}
where the input parameters can be found in \cite{Wu2006}.

\subsection{Helicity Amplitudes}
The helicity amplitude formalism is used in the analysis of the $c \rightarrow (s, d)\bar{\ell}\nu_\ell$ transitions. The relevant vector helicity amplitudes are given by \cite{Fleischer2020} 
\begin{eqnarray}
H^P_{V,t} &=& \frac{M^2_{D_{(s)}} - m_P^2}{\sqrt{q^2}}f_0(q^2)~, \nonumber \\
H^P_{V,0} &=& \sqrt{\frac{\lambda_P(q^2)}{q^2}}f_+(q^2)~,
\end{eqnarray} where 
\begin{equation}
\lambda_P(q^2) = [(M_{D_{(s)}}-m_P)^2-q^2][(M_{D_{(s)}}+m_P)^2-q^2].
\end{equation}

The scalar helicity amplitudes are obtained from the vector ones using equations of motion
\begin{equation}
H^P_{S} = \frac{M^2_{D_{(s)}}-m^2_P}{m_c-m_q}f_0(q^2)~.
\end{equation}

\subsection{$q^2$-dependent Observables}
The differential branching fraction is given by
	\begin{eqnarray}
			\frac{d\mathcal{B}}{dq^2} &=& \frac{G_F^2\vert V_{cq} \vert^2 \tau_D \sqrt{Q_+Q_-}}{256\pi^3M_D^3}\left(1-\frac{m_\ell^2}{q^2}\right) \nonumber \\ &\times&\bigg\{\frac{2}{3}\left[\vert1+C_{V_L}+C_{V_R}\vert^2 (\vert H^P_{V,0}\vert^2 + 3 \vert H^P_{V,t}\vert^2)\right]m_\ell^2 \nonumber \\ &+& 4 Re\bigg[ (C_{S_L} + C_{S_R}) (1 + 
			C_{V_L} + C_{V_R})^*\bigg] \nonumber \\ &\times& H^P_S H^P_{V,t} m_\ell \sqrt{q^2} +\bigg[2 (C_{S_L} + C_{S_R})^2 \vert H^P_S\vert^2 \nonumber \\ &+& \frac{4}{3} \vert 1 + C_{V_L} + C_{V_R}|^2 \vert H^P_{V,0}\vert^2\bigg] q^2 \bigg\}, 
		\end{eqnarray} where \begin{equation}
	Q_{\pm} = (M_{D_{(s)}}\pm m_P)^2 - q^2.
\end{equation}
For other $q^2$-dependent observables such as forward-backward asymmetry $A^\ell_{FB}(q^2)$, lepton polarization asymmetry $P^\ell_F(q^2)$ and convexity parameter $C^\ell_F(q^2)$, their definitions can be found in \cite{leng2020, sakaki2013, Becirevic2021}.

\section{Constraints on New Couplings}
	In our analysis, we constrain the parameter space of the scalar $C_S = C_{S_L} + C_{S_R}$ and vector $C_V = C_{V_L} + C_{V_R}$ new physics couplings using the experimental measurements of the branching fractions of semileptonic $D$ meson decays. For $c \rightarrow s$ transitions, we use the ratio of branching fractions, based on same quark transition, to constrain $C_S$. This ratio $\mathcal{R}$ is defined as
\begin{equation}
\mathcal{R} \equiv \frac{\mathcal{B}(D_s^+ \rightarrow \eta\mu^+ \nu_\mu)}{\mathcal{B}(D_s^+ \rightarrow \eta^{\prime}\mu^+ \nu_\mu)}. 
\end{equation} For the vector coupling, the NP-dependence cancels out in the ratio and hence, for $C_V$, we obtain constraints using $\mathcal{B}(D_s^+ \to \eta^{(\prime)} \mu^+ \nu_\mu)$ only.
For $c \rightarrow d$ transitions, due to the unavailability of the experimental measurement of $\mathcal{B}(D^+\rightarrow\eta^{\prime} \mu^+\nu_\mu)$, we only use the $\mathcal{B}(D^+\rightarrow\eta\mu^+\nu_\mu)$ data to constrain the couplings.  The currently available experimental measurements \cite{Ablikim2017,Ablikim2020} of the branching fractions are given in Table \ref{2_table}.
\begin{table}[h]
	\caption{Current experimental results for branching fractions}
\begin{center}
	\begin{tabular}{|c|c|}
		\hline
		Decay & Experiment\\
		\hline
		$\mathcal{B}(D^+\rightarrow\eta\mu^+\nu_\mu)$ & $(1.04\pm0.11)\times 10^{-3}$\\
		\hline
		$\mathcal{B}(D_s^+\rightarrow\eta\mu^+\nu_\mu)$ & $(2.4\pm0.5)\times 10^{-2}$\\
		\hline
		$\mathcal{B}(D_s^+\rightarrow\eta'\mu^+\nu_\mu)$ & $(11.0\pm5.0)\times 10^{-3}$\\
		\hline
	\end{tabular} \label{2_table}
\end{center} 
\end{table}

\section{Results and Discussion}
\begin{figure}[h]
	\centering
		\includegraphics[scale=0.34]{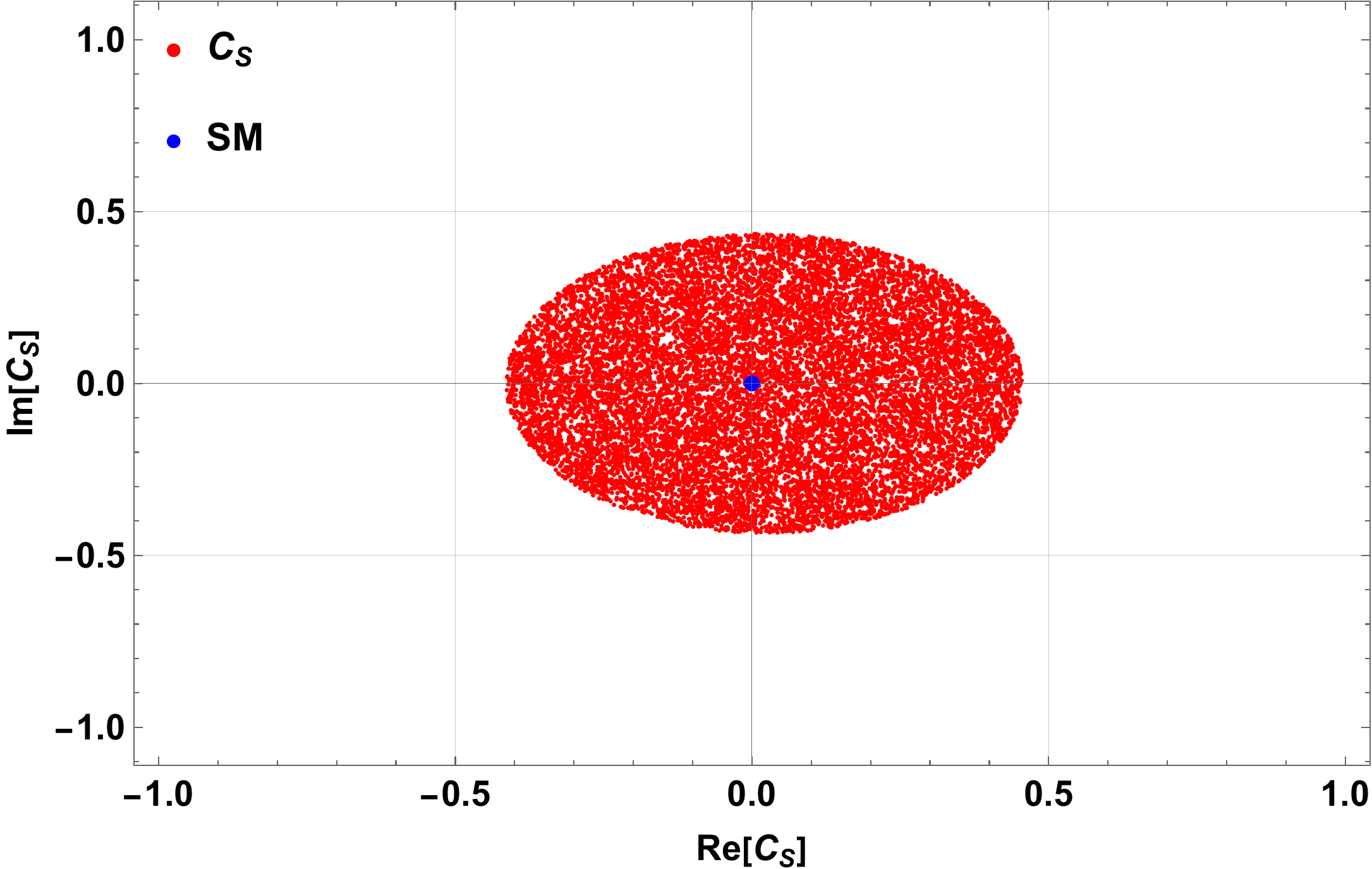}
		\caption{{\small The allowed parameter space of the scalar coupling $C_S$ using $\mathcal{R}$ for $c \rightarrow s$ transitions.}} \label{fig1}
\end{figure}

\begin{figure}
\centering
\includegraphics[scale=0.34]{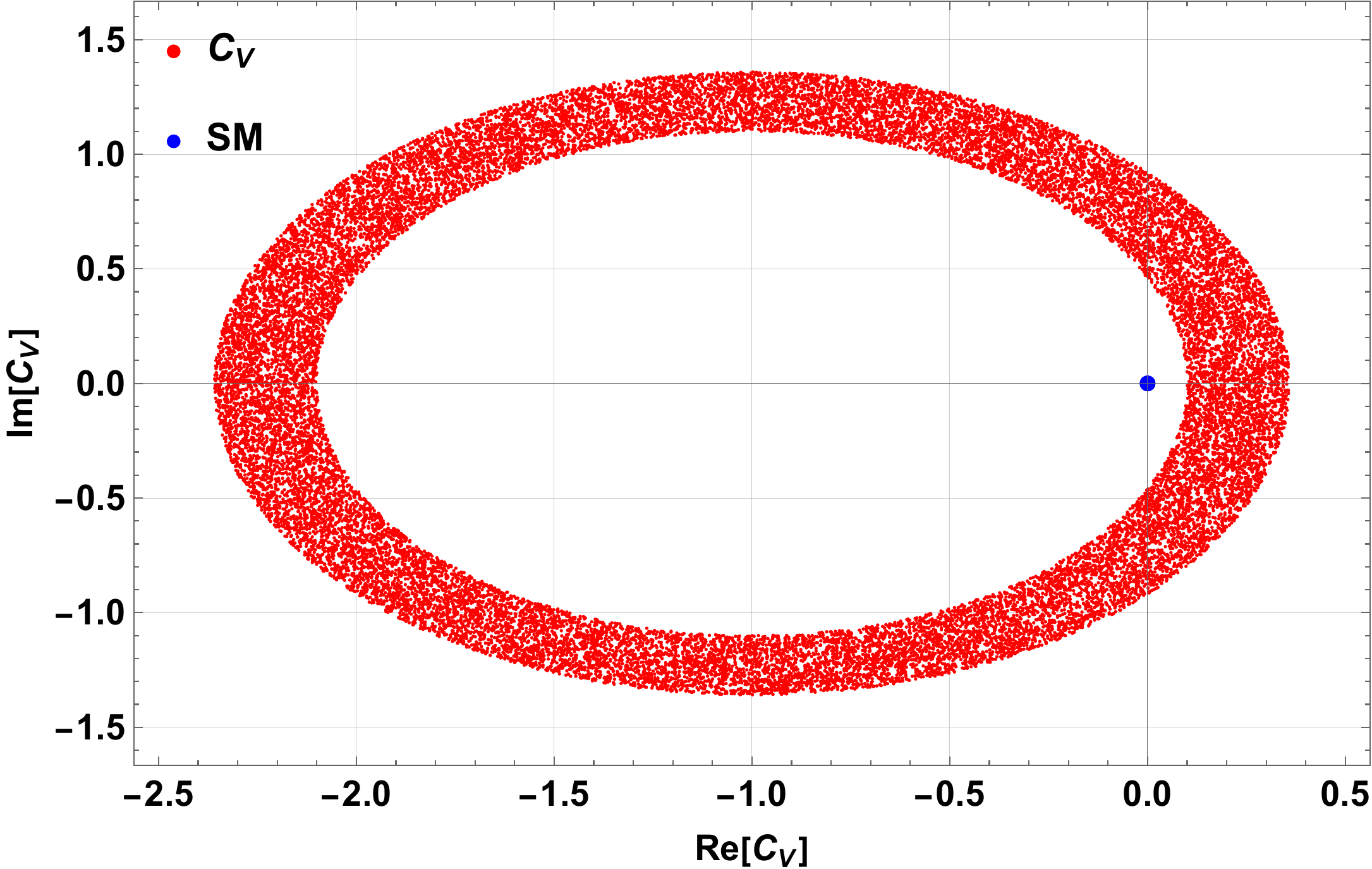}
\caption{{\small The allowed parameter space of the vector coupling $C_V$ using $\mathcal{B}(D_s^+\rightarrow\eta^{(\prime)}{\mu^+}\nu_\mu)$ for $c \rightarrow s$ transitions.}} \label{fig2}
\end{figure}

\begin{figure}
\centering
\includegraphics[scale=0.34]{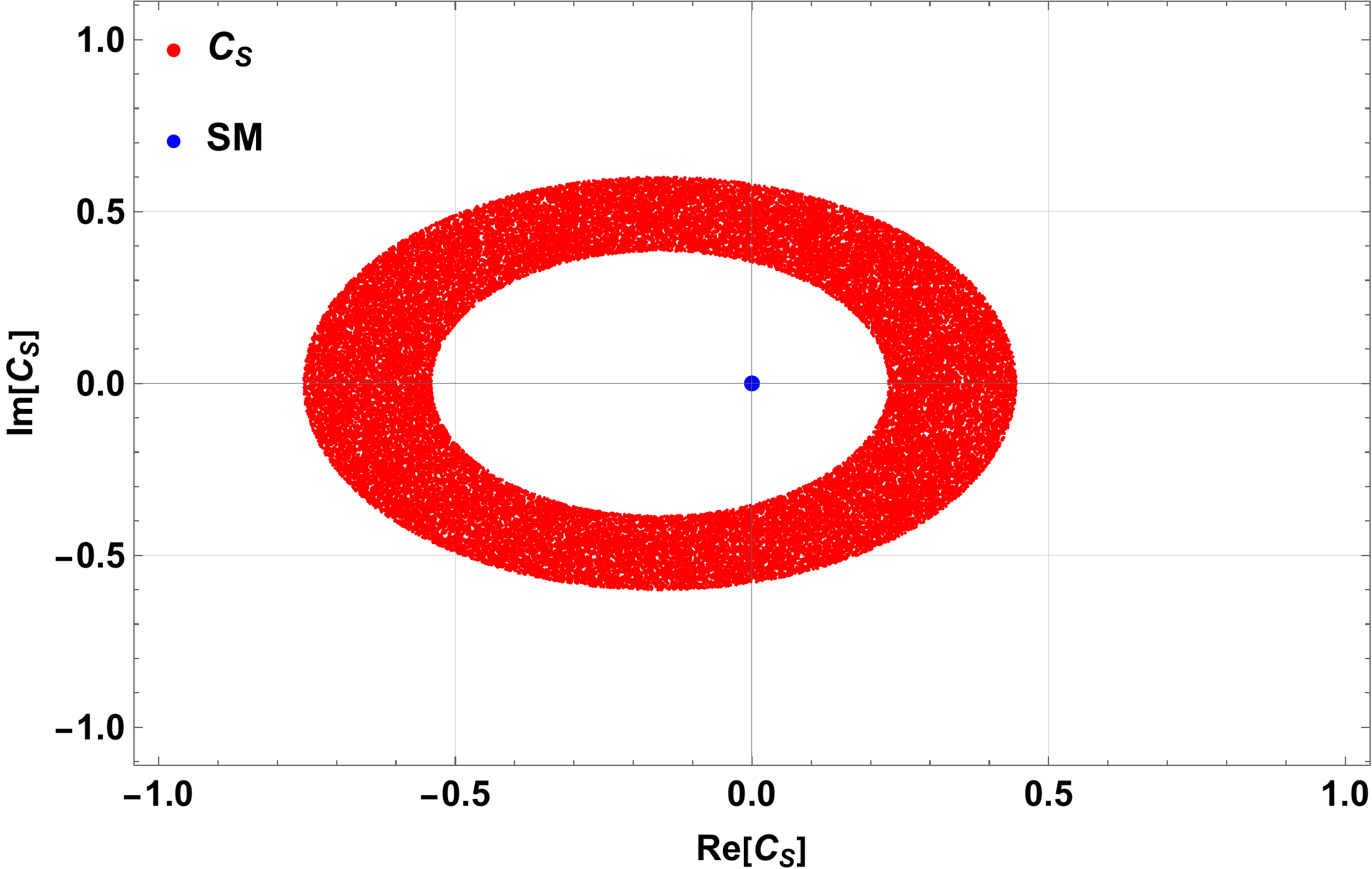}
\caption{\small{The allowed parameter space of the scalar coupling $C_S$ using $\mathcal{B}(D^+\rightarrow\eta\bar{\ell}\nu_{\ell})$ for $c \rightarrow d$ transitions.}} \label{fig3}
\end{figure}

\begin{figure}
\centering
\includegraphics[scale=0.34]{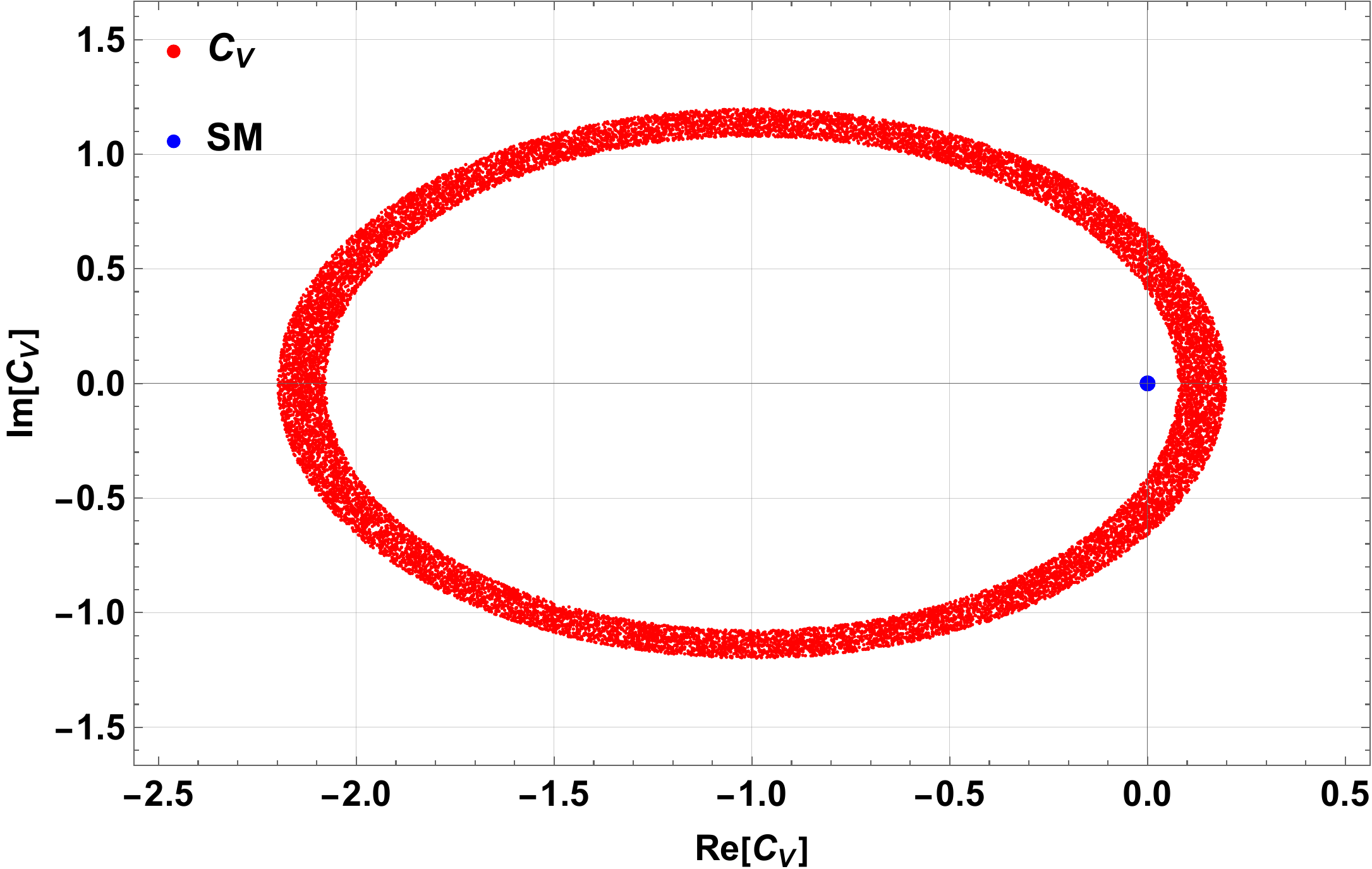}
\caption{\small{The allowed parameter space of the vector coupling $C_V$ using $\mathcal{B}(D^+\rightarrow\eta\bar{\ell}\nu_{\ell})$ $c \rightarrow d$ transitions.}} \label{fig4}
\end{figure}

The obtained constraint parameter space for $C_S$ and $C_V$ couplings ($c \rightarrow s$ transitions) are given in Fig. (\ref{fig1}) and Fig. (\ref{fig2}), respectively. Figures (\ref{fig3}) and (\ref{fig4}) are the respective allowed parameter space for $C_S$ and $C_V$ couplings ($c \rightarrow d$ transitions). 

\begin{figure}
	\centering
			\includegraphics[scale=0.17]{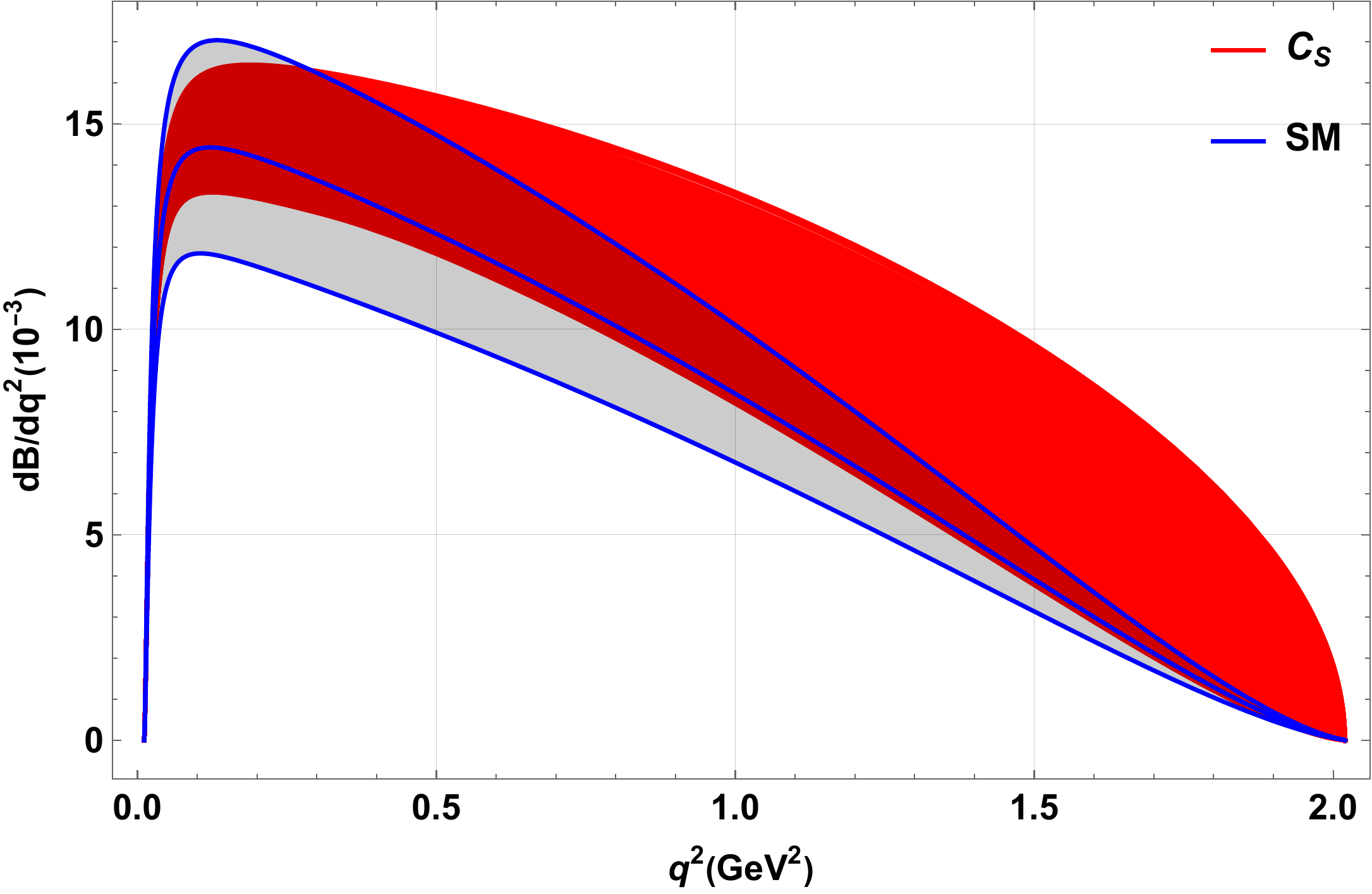}
			\includegraphics[scale=0.17]{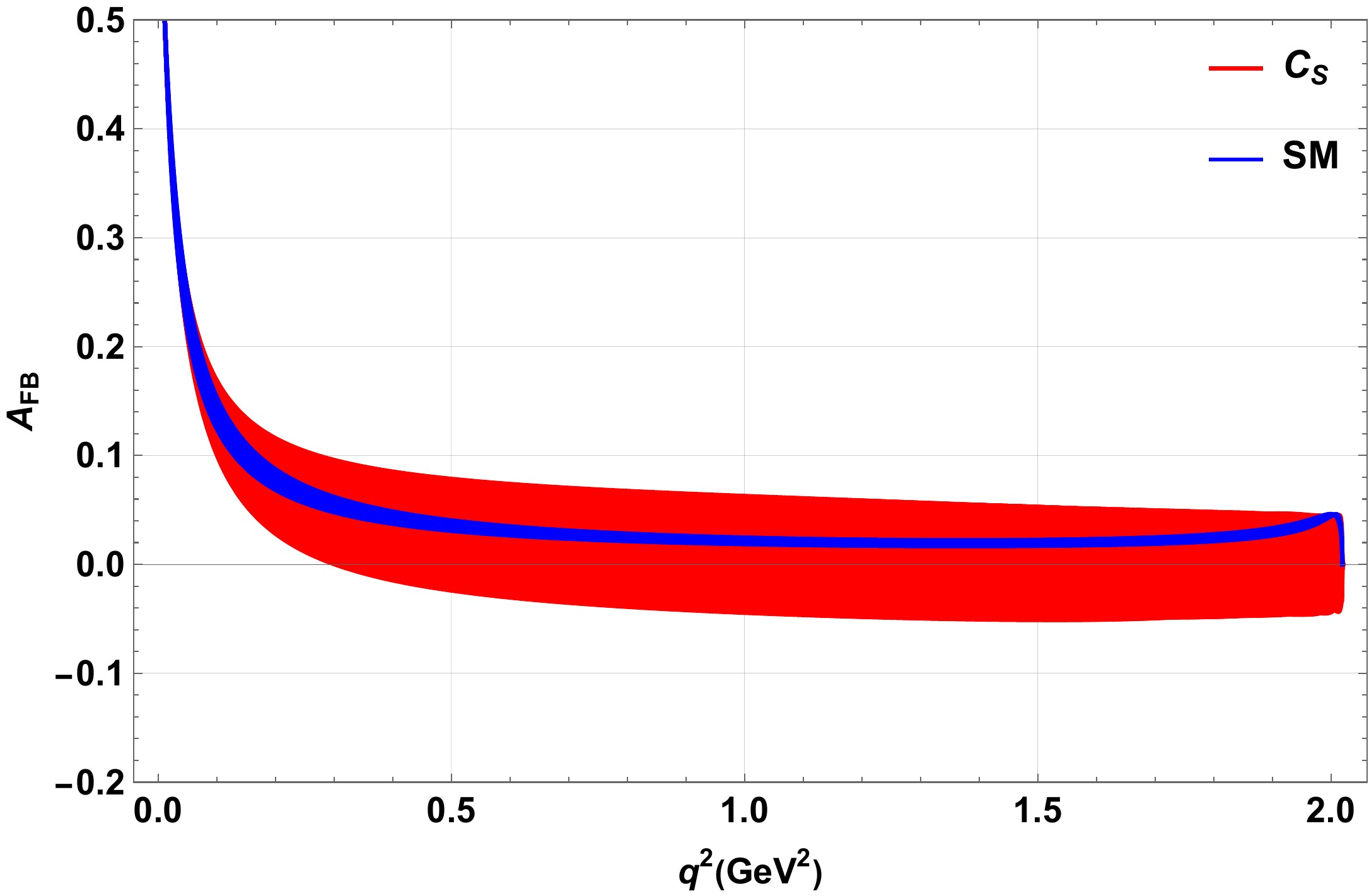}
			\includegraphics[scale=0.17]{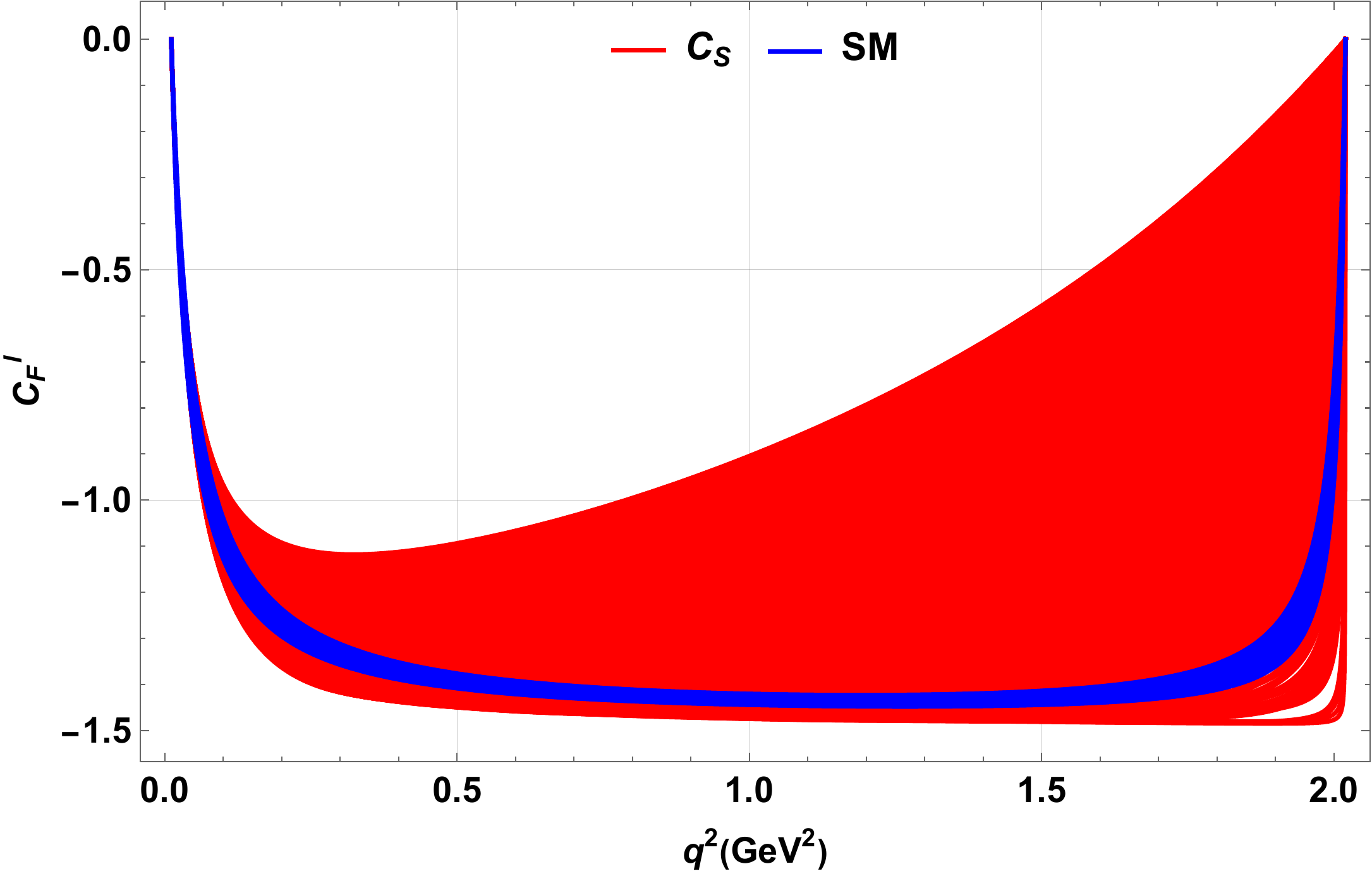}
			\includegraphics[scale=0.17]{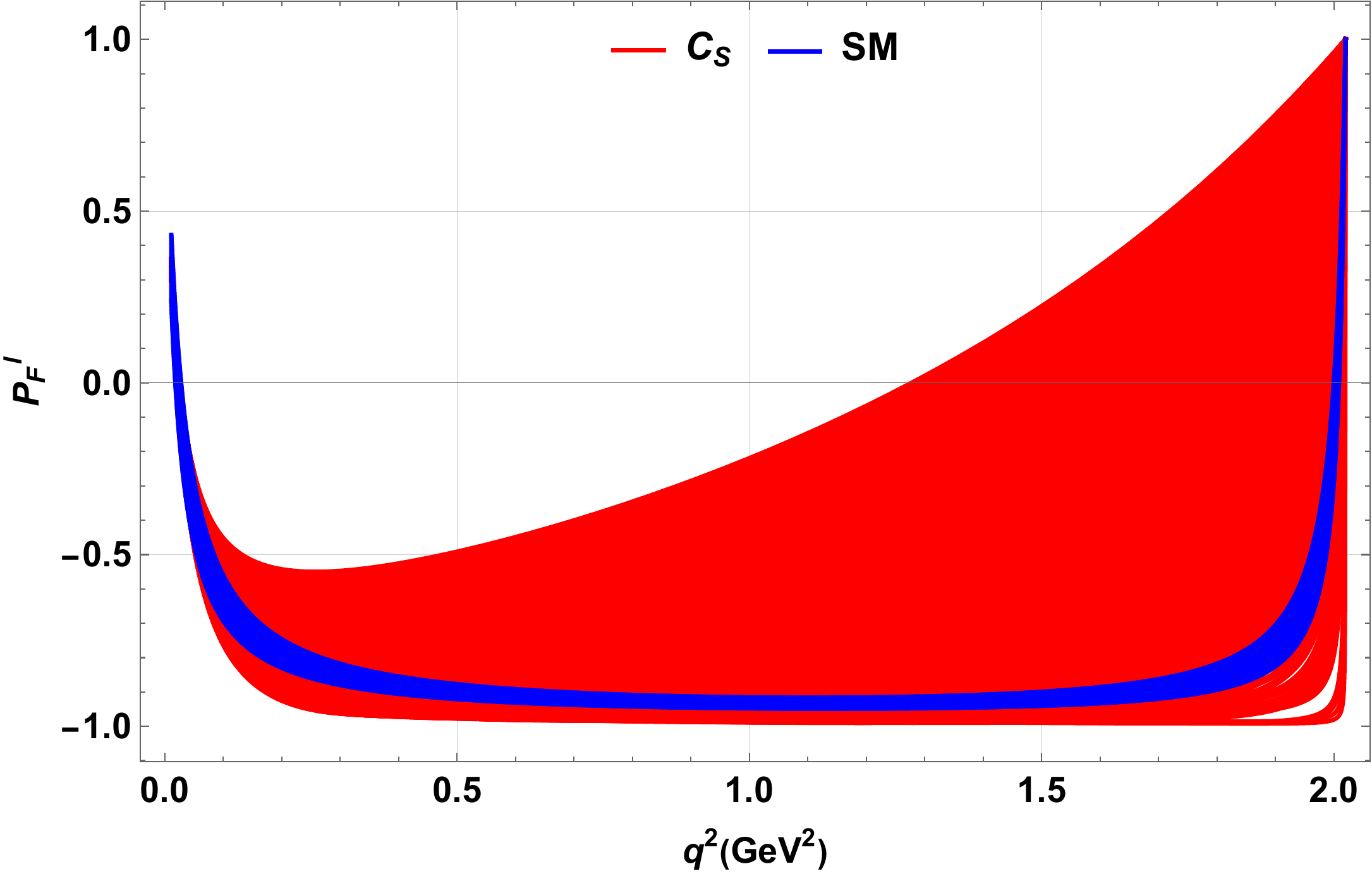}
		\caption{\small{$q^2$-dependence of various observables for $D_{s}^+\rightarrow \eta\mu^+\nu_\mu$  in presence of $C_S$}} \label{fig5}
\end{figure}

\begin{figure}
\centering
	\includegraphics[scale=0.17]{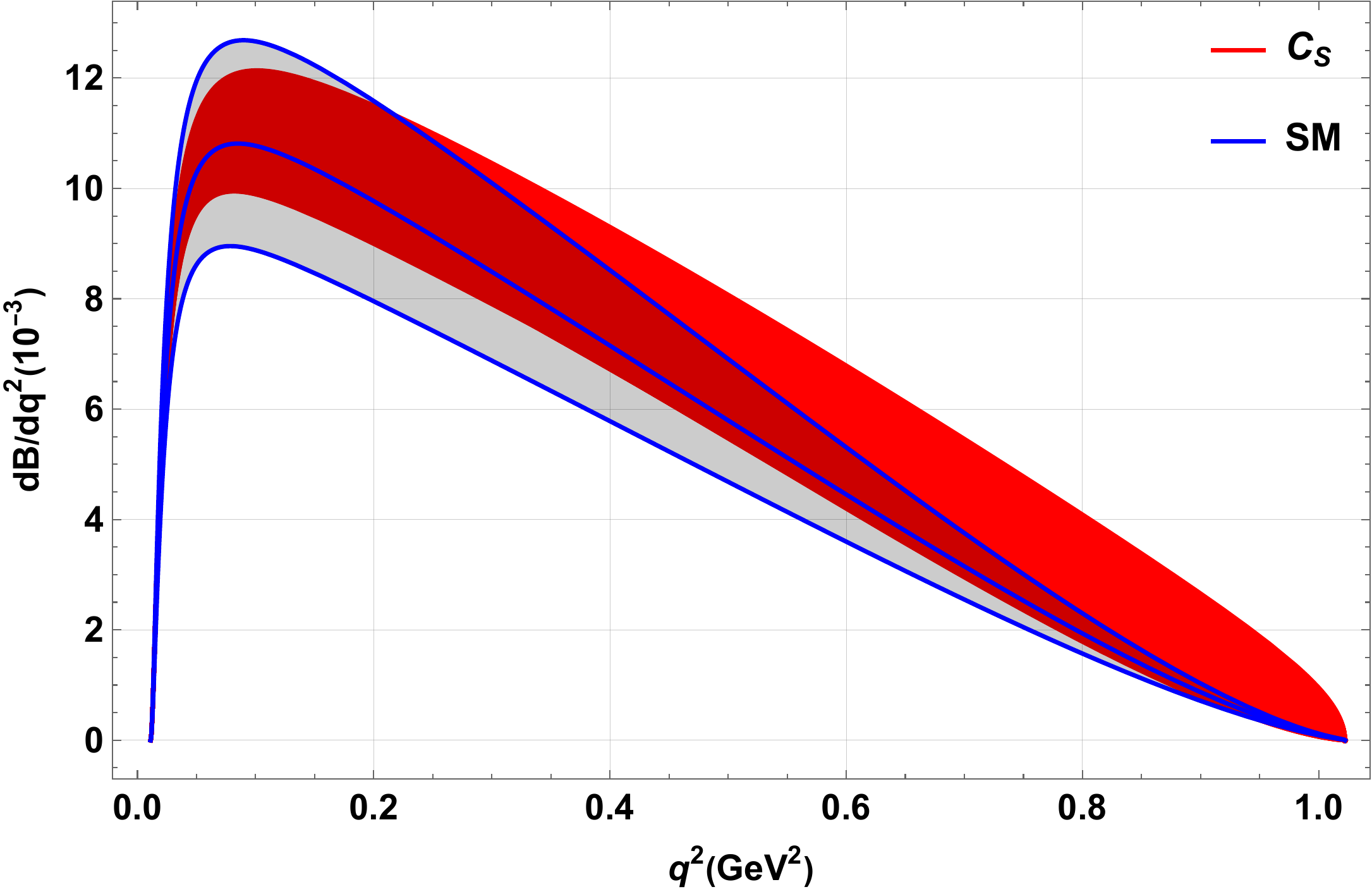}
	\includegraphics[scale=0.17]{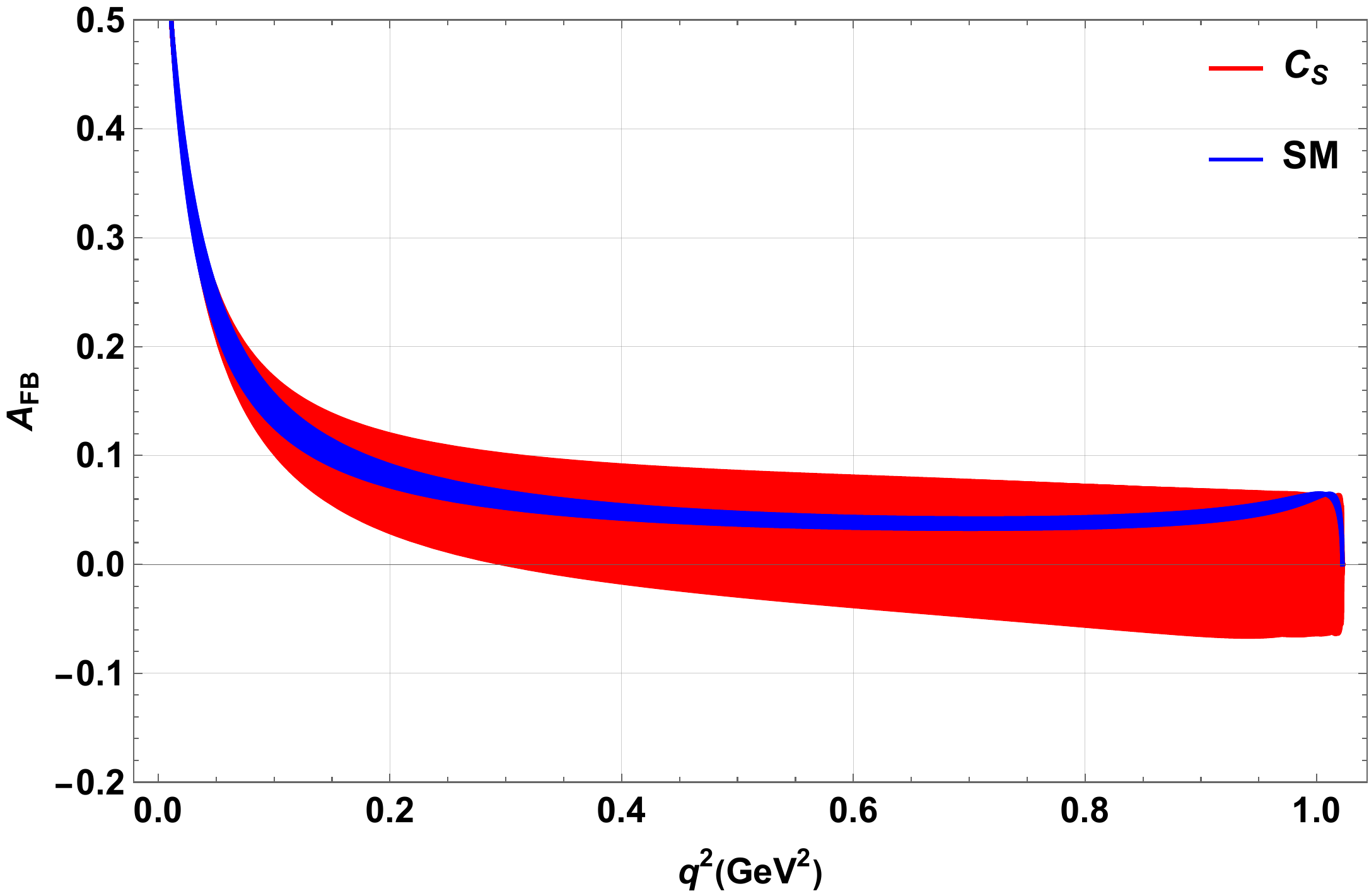}
	\includegraphics[scale=0.17]{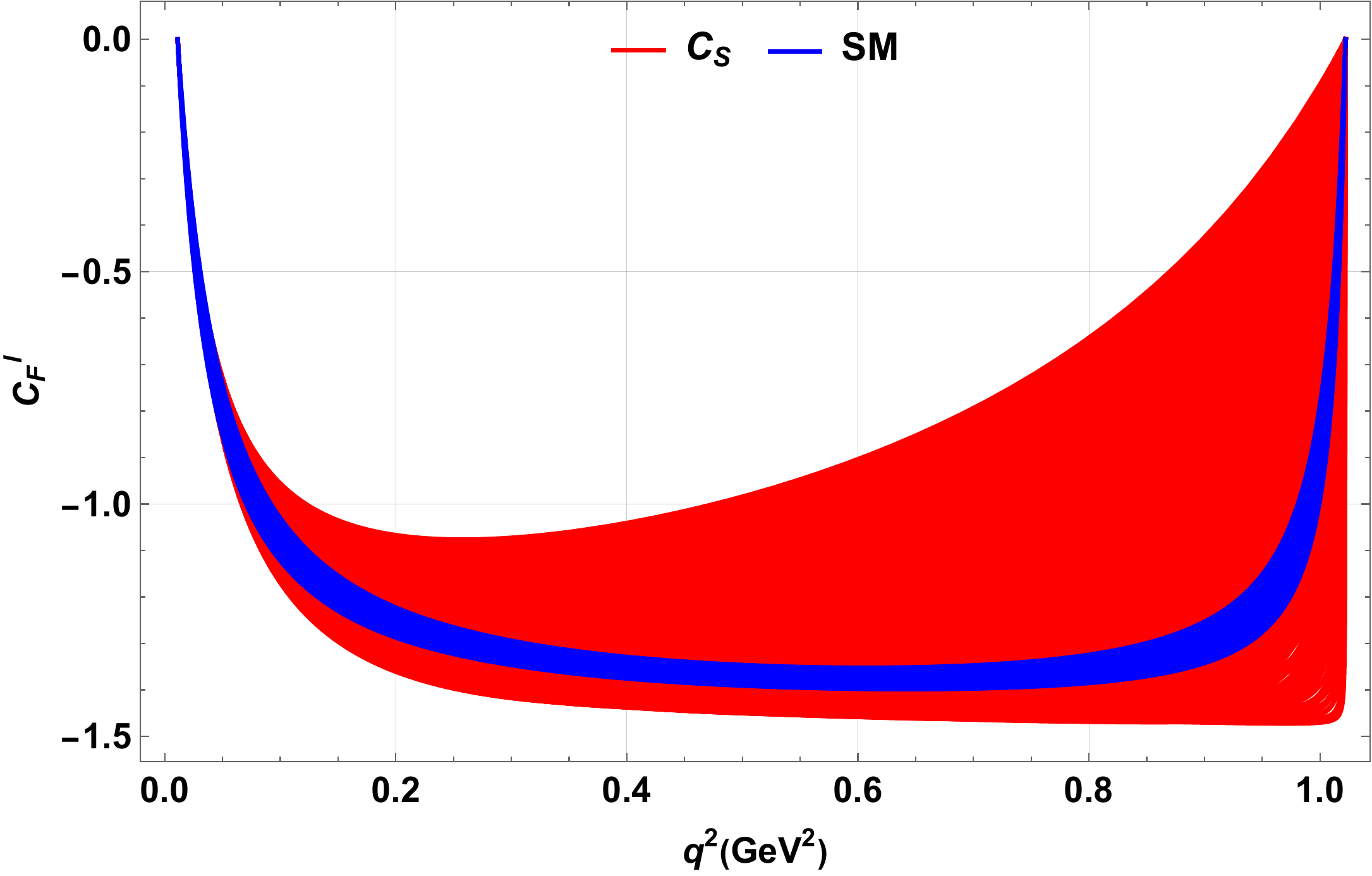}
	\includegraphics[scale=0.17]{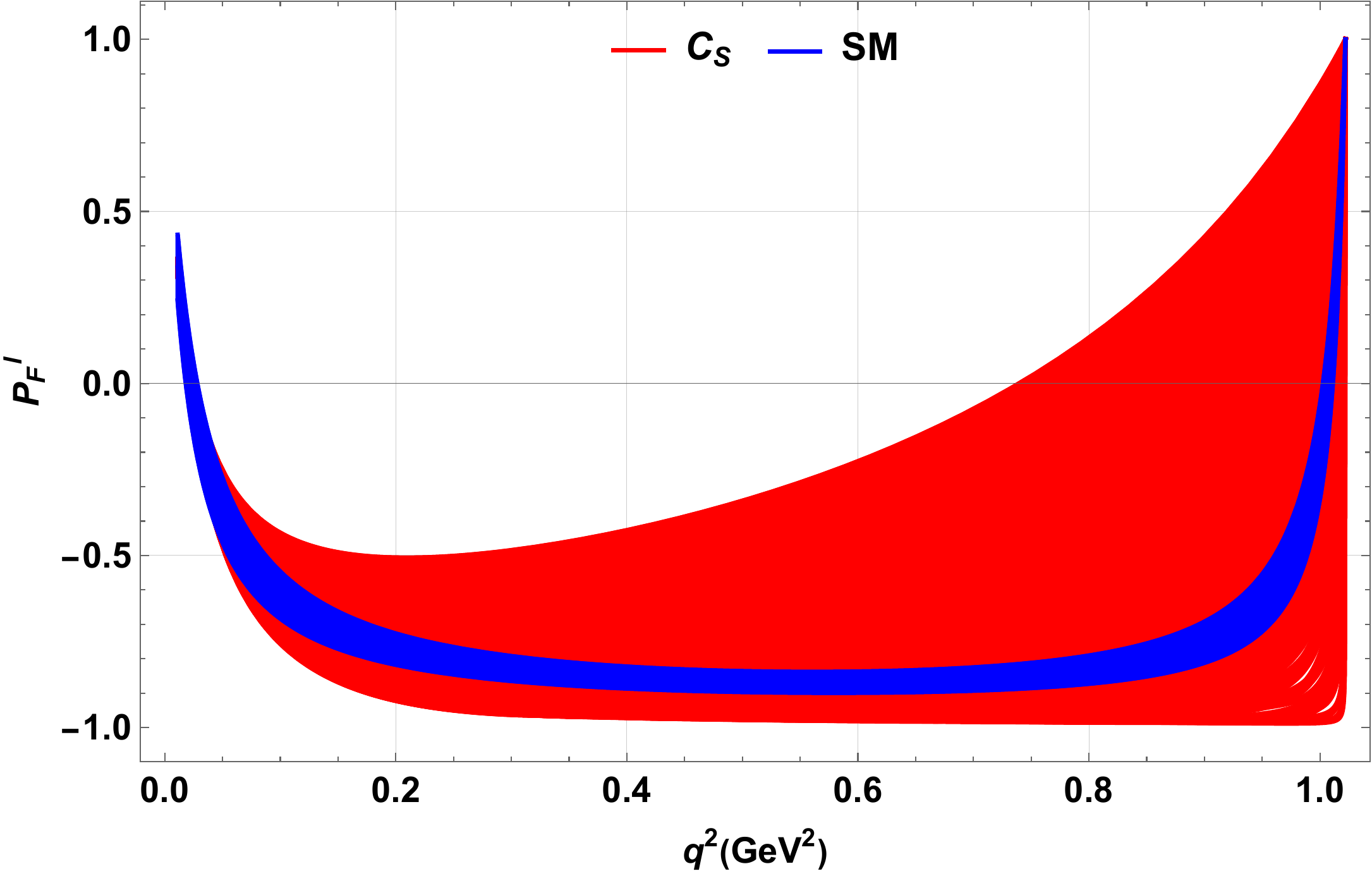} 
	\caption{\small{$q^2$-dependence of various observables for $D_{s}^+\rightarrow \eta^{\prime} \mu^+ \nu_\mu$ in presence of $C_S$.}} \label{fig6}
\end{figure}

For $c\rightarrow s$ transitions, the $q^2$-dependence of various observables are shown in Fig. (\ref{fig5}) and Fig. (\ref{fig6}) in the presence of $C_S$. We observe that NP effects are more pronounced in $P^\ell_F(q^2)$ and $C^\ell_F(q^2)$, as compared to $\frac{d\mathcal{B}}{dq^2} $ and $A^\ell_{FB}(q^2)$. There is also a maximal cancellation of theoretical uncertainties in the SM for the ratioed observables. 

\begin{figure}
	\centering
		\includegraphics[scale=0.17]{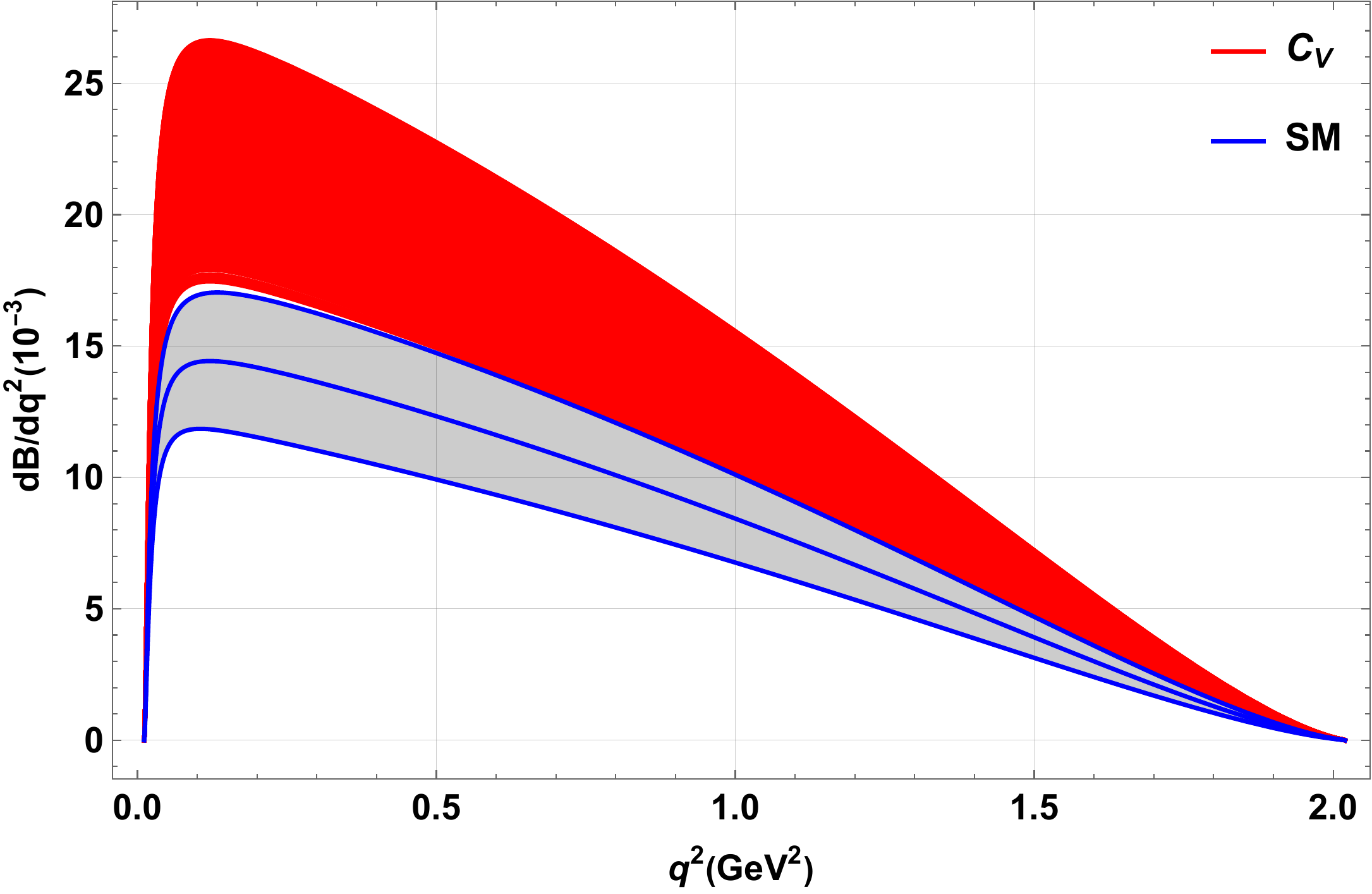}
		\includegraphics[scale=0.17]{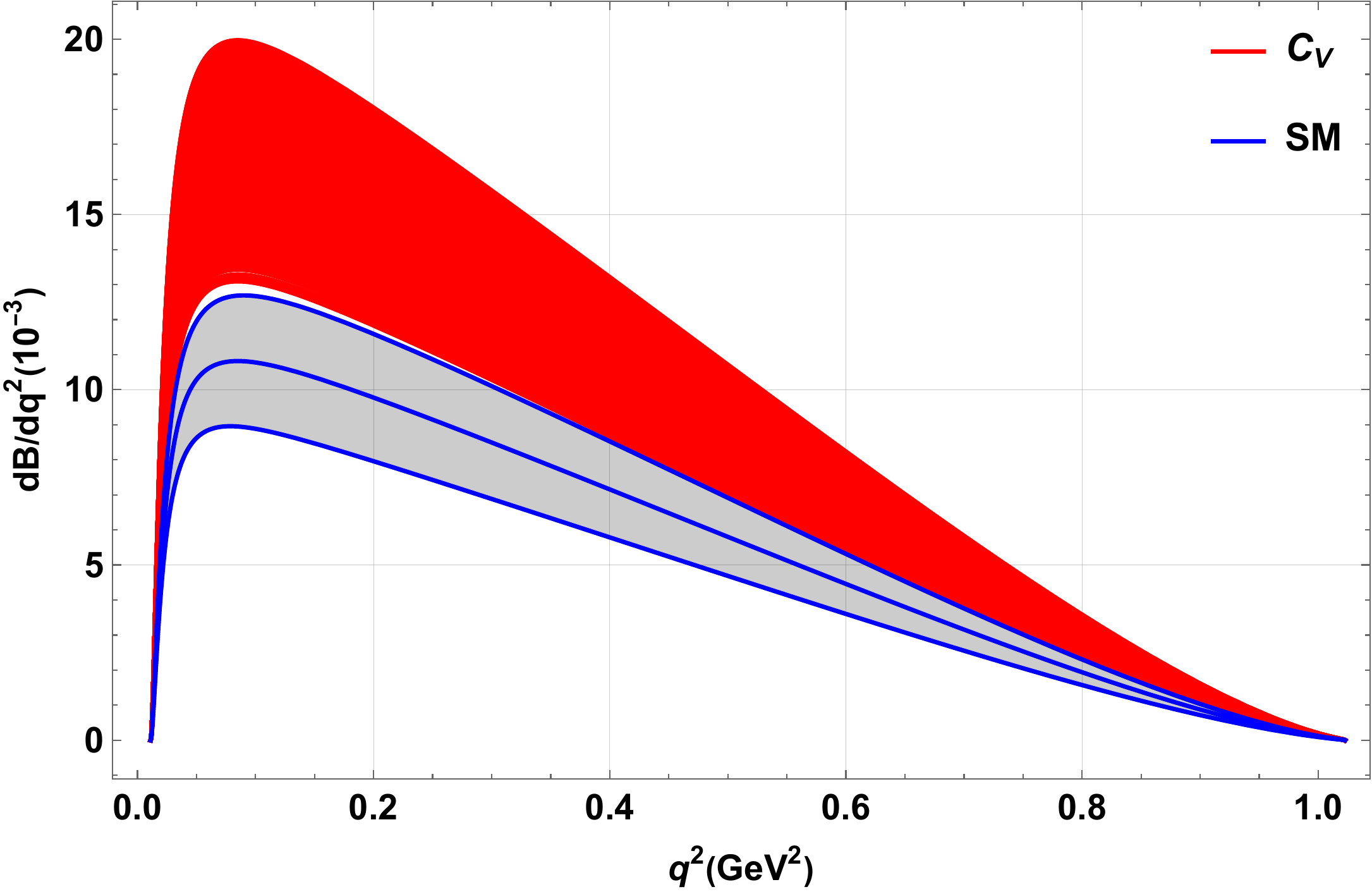}
		\caption{\small{$q^2$-dependence of $\frac{d\mathcal{B}}{dq^2} $ for $D_{s}^+\rightarrow \eta \mu^+\nu_\mu$ (left panel) and $D_{s}^+\rightarrow \eta^{\prime}\mu^+\nu_\mu$ (right panel) in presence of $C_V$.}} \label{fig7}
\end{figure}

In the case of the vector coupling $C_V$, the NP-dependence of all the observables cancels out, except for the differential branching fraction, the plot of which is shown in Fig. (\ref{fig7}). For $\frac{d\mathcal{B}}{dq^2} $, the NP effects are observed to be slightly higher in the presence of $C_V$ as compared to the scalar one.

\begin{figure}[h]
	\centering
			\includegraphics[scale=0.17]{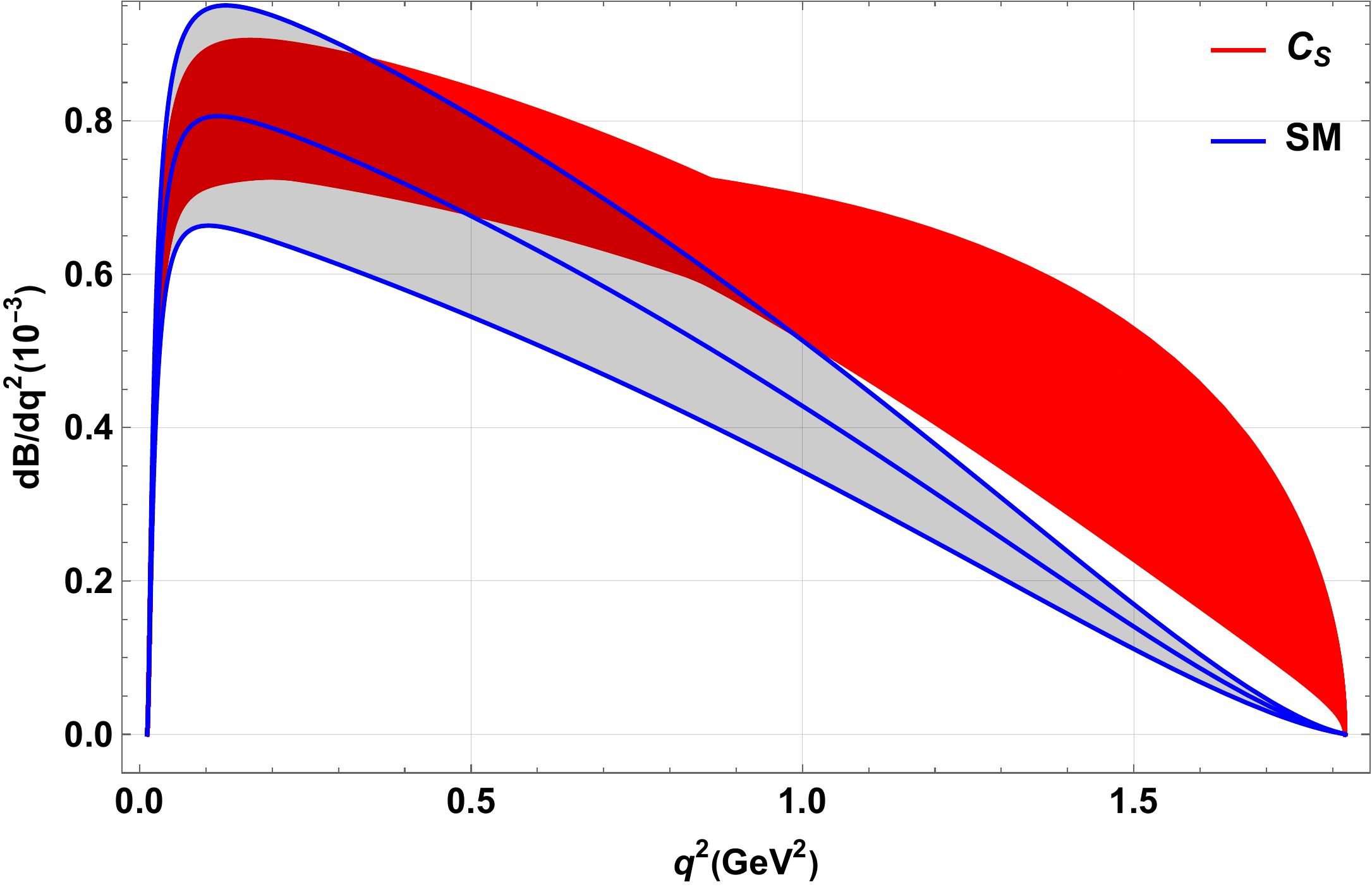}
			\includegraphics[scale=0.17]{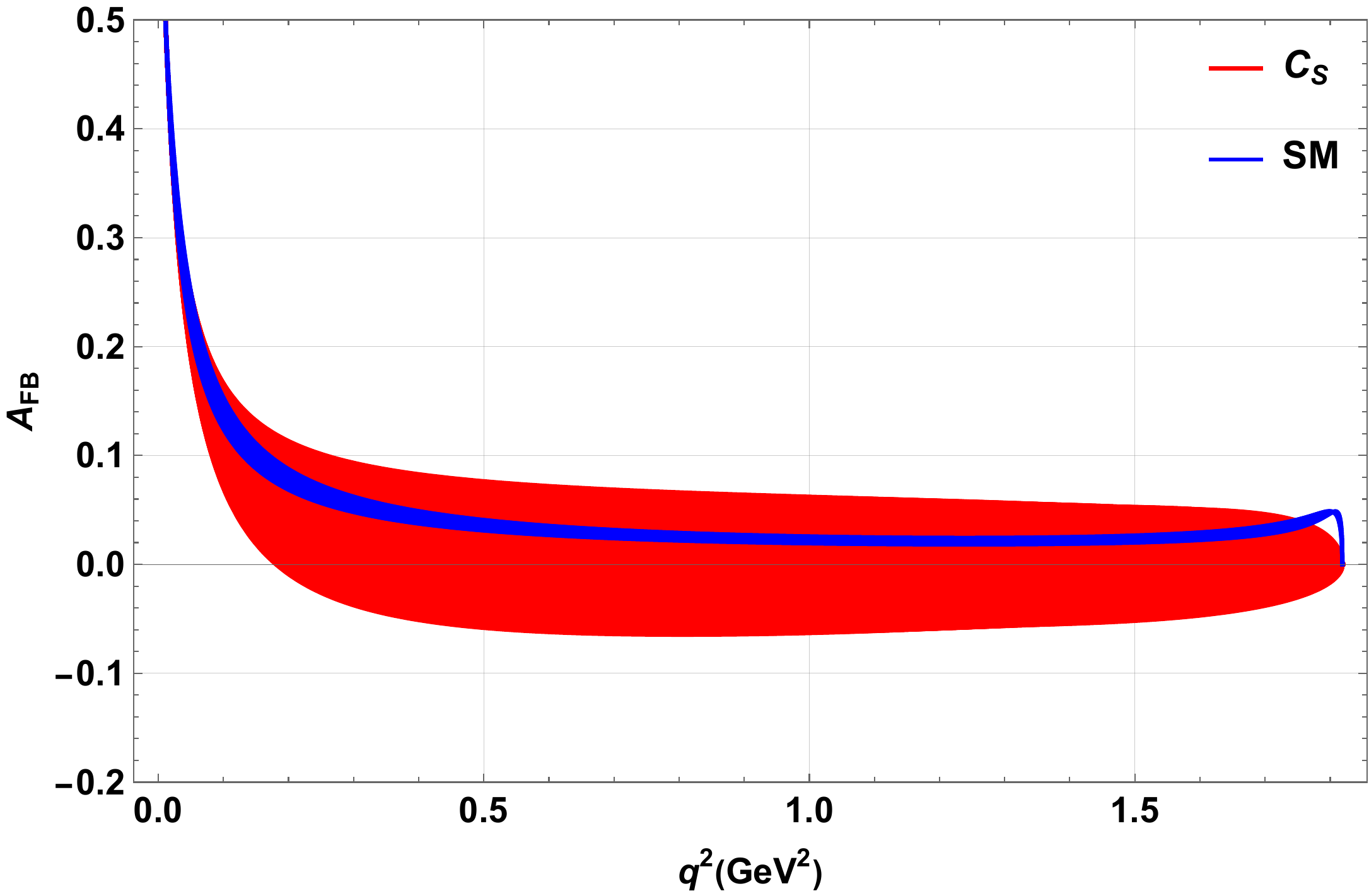}
			\includegraphics[scale=0.17]{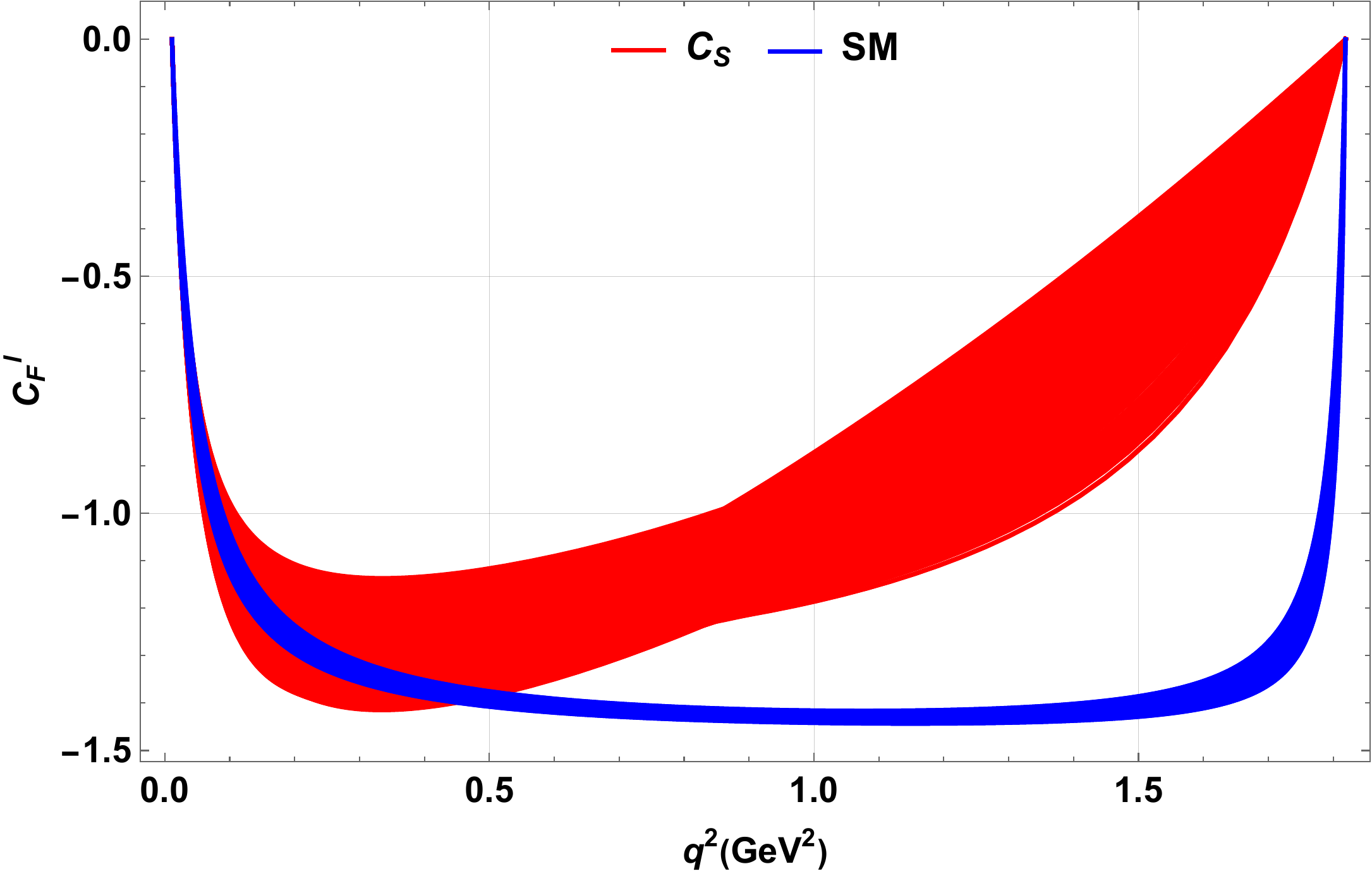}
			\includegraphics[scale=0.17]{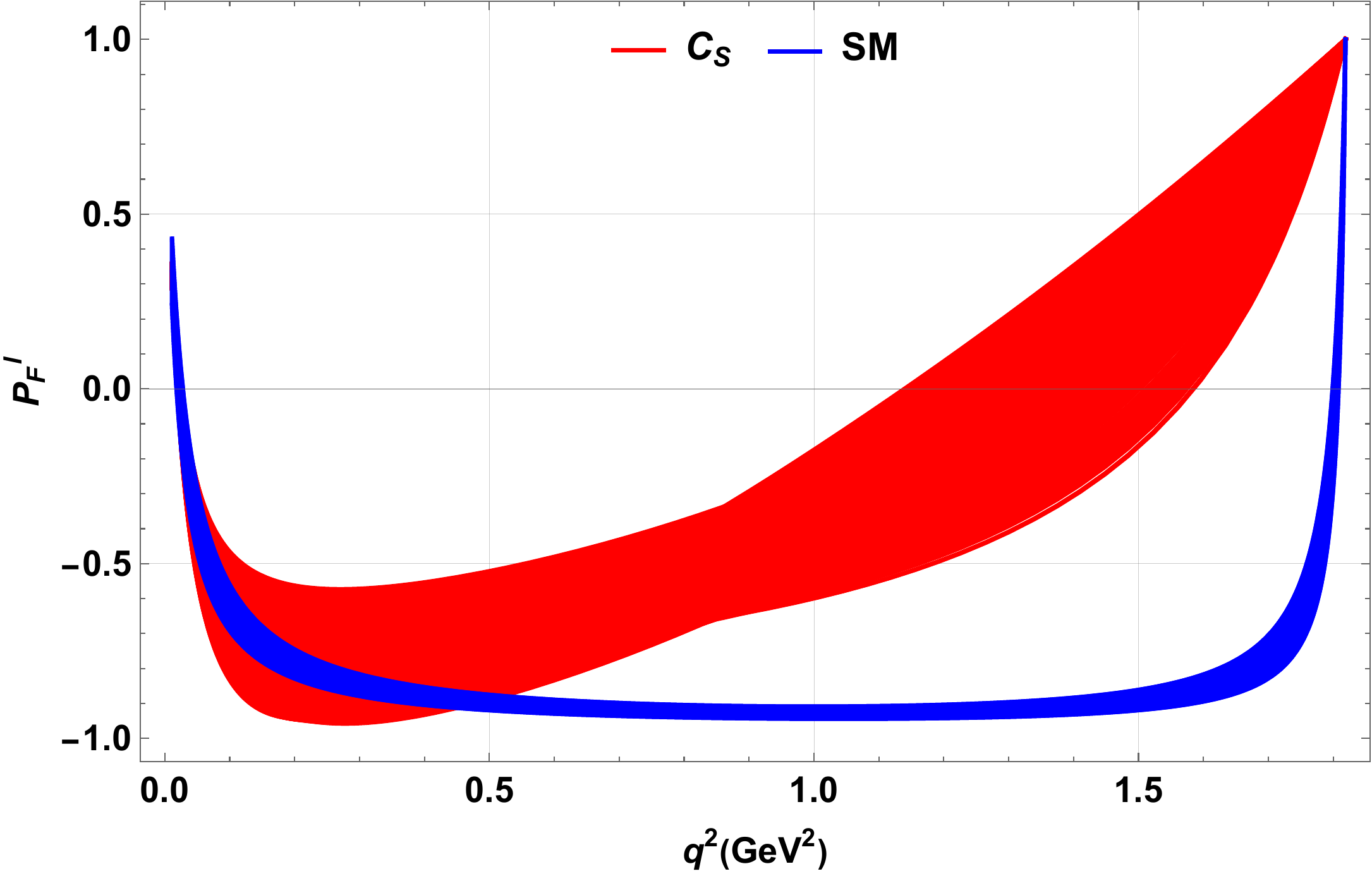}
	\caption{\small{$q^2$-dependence of various observables for  $D^+\rightarrow \eta\mu^+\nu_\mu$  in presence of $C_S$.}} \label{fig8}
\end{figure}

\begin{figure}[h]
	\centering
		\includegraphics[scale=0.34]{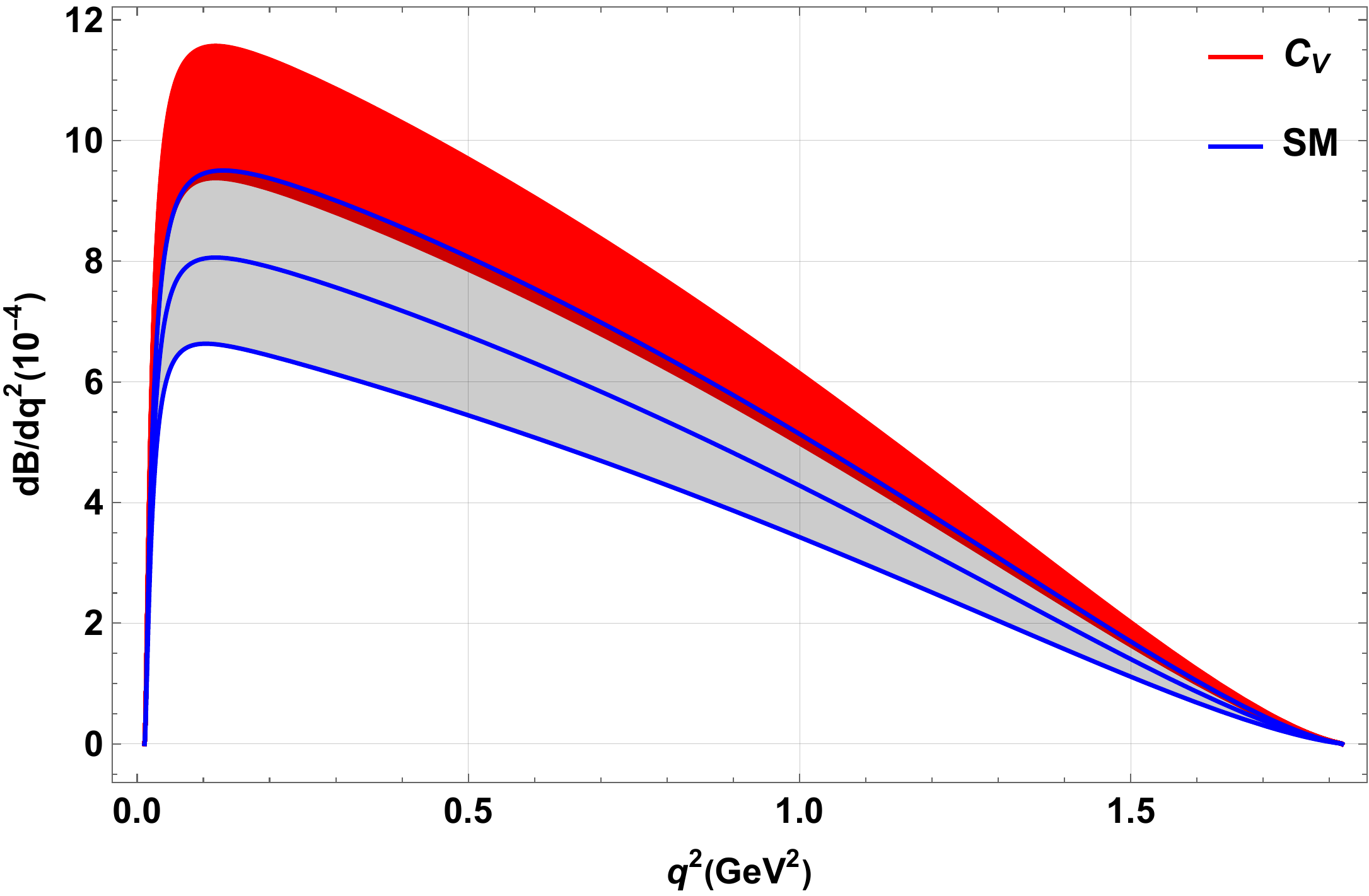}
		\caption{\small{$q^2$-dependence of differential branching fraction for $D^+\rightarrow \eta\mu^+\nu_\mu$  in presence of $C_V$.}} \label{fig9}
\end{figure}

For $c\rightarrow d$ transitions, the $q^2$-dependence of various observables are shown in Fig. (\ref{fig8}) and Fig. (\ref{fig9}) in the presence of $C_S$ and $C_V$, respectively. Even in this case, we observe $P^\ell_F(q^2)$ and $C^\ell_F(q^2)$ display more deviation from the SM prediction as compared to $\frac{d\mathcal{B}}{dq^2} $ and $A^\ell_{FB}(q^2)$ for $C_S$. Also, NP effects on $\frac{d\mathcal{B}}{dq^2} $ can be differentiated slightly more than for $c\rightarrow s$ case.

\section{Conclusion}
We have analyzed the decay modes $D^+_{(s)} \rightarrow \eta^{(\prime)}\mu^+\nu_\mu$ within the SM as well as beyond, using an effective Lagrangian approach. We used the available experimental measurements of semileptonic $D$ meson decays to constrain the parameter space of NP couplings. We probed the NP dependence of various observables and noticed that in the presence of scalar couplings, the NP sensitivity was more pronounced in $P^\ell_F(q^2)$ and $C^\ell_F(q^2)$. In case of vector couplings, the NP-dependence cancelled out in all observables, except in the branching fraction. Studies of charm meson decays as those in this work provide a unique opportunity to probe NP beyond SM in the up-sector. We hope that future measurements of new modes and observables, and improved measurements of existing modes will help in obtaining stronger constraints on possible NP contributions. 

\bigskip 

\end{document}